\documentclass[aps,prl,twocolumn,superscriptaddress,amsfont,graphicx,nofootinbib,preprintnumbers]{revtex4}
\usepackage{dcolumn}
\usepackage{amsmath}
\usepackage{amsthm}
\usepackage{amstext}
\usepackage{amsfonts}
\usepackage{xcolor}
\usepackage{graphicx}
\usepackage{hyperref}
\usepackage{empheq}
\usepackage{amssymb}

\definecolor{nicered}{rgb}{0.7,0.1,0.1}
\definecolor{nicegreen}{rgb}{0.0,0.4,0.0}
\hypersetup{colorlinks,citecolor=nicegreen,linkcolor=nicered,urlcolor=nicered}

\def\nn{\notag}

\definecolor{lightgreen}{rgb}{0.56, 0.69, 0.19}
\definecolor{lightblue}{rgb}{0.36, 0.51, 0.71}
\definecolor{lightyellow}{rgb}{0.88, 0.61, 0.14}
\definecolor{darkgreen}{rgb}{0.6, 0.6, 0.35}
\definecolor{lightred}{rgb}{0.99, 0.36, 0.02}
\definecolor{box1}{rgb}{0.46, 0.6, 0.45}
\definecolor{box2}{rgb}{0.62, 0.56, 0.43}
\definecolor{box3}{rgb}{0.72, 0.65, 0.17}
\definecolor{myorange}{RGB}{230,120,20}

\definecolor{darkred}{rgb}{0.5,0.0,0.0}
\definecolor{darkblue}{rgb}{0.0,0.0,0.9}
\definecolor{darkerblue}{rgb}{0.0,0.0,0.5}
\definecolor{darkgreen}{rgb}{0.0,0.5,0.0}
\definecolor{black}{rgb}{0.0,0.0,0.0}
\definecolor{brown}{rgb}{0.6,0.4,0.2}

\newcommand{\cin}[1]{{\color{darkred}{#1}}}
\newcommand{\ccb}[1]{{\color{lightblue}{#1}}}
\newcommand{\cca}[1]{{\color{darkerblue}{#1}}}

\newcommand{\cce}[1]{{\color{myorange}{#1}}}

\begin{document}

\def\CERN{CERN, Theoretical Physics Department, CH-1211 Geneva 23, Switzerland}
\def\SJTU{School of Physics and Astronomy, Shanghai Jiao Tong University, Shanghai 200240, China}
\def\KLPAC{Key Laboratory for Particle Astrophysics and Cosmology (MOE), Shanghai 200240, China }

\preprint{CERN-TH-2026-066}

\title{Two-Loop Spacelike Splitting Amplitudes in Full-Color QCD}

\author{Federico Buccioni}            
\email[Electronic address: ]{federico.buccioni@cern.ch}
\affiliation{\CERN}

\author{Hanyu Fang}
\email[Electronic address: ]{aurora\_superposition@sjtu.edu.cn}
\affiliation{\SJTU}
\affiliation{\KLPAC}

\author{Kai Yan}         
\email[Electronic address: ]{yan.kai@sjtu.edu.cn}
\affiliation{\SJTU}
\affiliation{\KLPAC}


\begin{abstract}
The study of QCD scattering amplitudes in the collinear regime provides crucial insight into the factorization properties of hadronic cross sections.
In this paper, we present the first complete results for two-loop spacelike splitting amplitudes in full-color QCD, in all partonic channels and helicity configurations.
We confirm the universality of a class of contributions already found in $\mathcal{N}=4$ super Yang–Mills (sYM) theory, and identify previously unknown sources of collinear factorization-violating (CFV) effects.
Consistent with recent observations in $\mathcal{N}=4$ sYM, all CFV contributions cancel in color-summed squared amplitudes, implying the universality of single-parton collinear factorization for jet cross sections at third order in QCD.
\end{abstract}

\maketitle

\section{Introduction}
\label{se:introduction}
Factorization in Quantum Chromodynamics (QCD) is central to theoretical predictions for hadron-collider cross sections, as it separates the long-distance hadron dynamics from the short-distance hard scattering. This is formalized by the factorization formula~\cite{Collins:1985ue,Collins:1989gx,Collins:1988ig}
\begin{equation}
\label{eq:hadronicxs}
\mathrm{d}\sigma_{h_1 h_2\to X} = \sum_{ij}\int\mathrm{d}x\mathrm{d}y ,f_{i/h_1}(x)f_{j/h_2}(y)\,\mathrm{d}\hat{\sigma}_{ij\to X},
\end{equation}
which expresses the cross section for the production of a final state $X$ in the scattering of hadrons $h_{1,2}$ as the convolution of non-perturbative parton distribution functions (PDFs) $f_{i/h_i}(x)$ and a perturbatively calculable partonic cross section $\mathrm{d}\hat{\sigma}$. 
Collinear factorization and its universality lie at the heart of Eq.~\eqref{eq:hadronicxs}, implying that the separation of long- and short-distance dynamics is independent of the final state. Violations of this universality would severely compromise both its predictive power and the interpretation of collider data within QCD.

Violations of collinear factorization in jet cross sections have been extensively studied~\cite{Bomhof:2004aw,Bacchetta:2005rm,Bomhof:2006dp,Collins:2007nk,Forshaw:2008cq,Catani:2011st,Forshaw:2012bi,Gaunt:2014ska,Zeng:2015iba}. These effects arise from loop corrections mediated by soft (Glauber) gluons, which break color coherence beyond the planar limit of QCD~\cite{Banfi:2010xy,Forshaw:2021fxs}.
In spacelike collinear splittings, where one or more collinear particles are emitted from an incoming parton, Glauber gluons induce a subleading $1/N_c$ dependence
on the quantum numbers of non-collinear partons~\cite{Catani:2011st,Cieri:2024ytf}, thereby violating \textit{strict collinear factorization}.

Coherence-violating Glauber phases in the hard amplitude have important phenomenological consequences. Among others, they are responsible for the appearance of so-called super-leading logarithms in jet cross sections~\cite{Forshaw:2008cq,Banfi:2010xy,Forshaw:2012bi,Becher:2024nqc}. Recently, their manifestation in global and non-global observables has attracted significant attention~\cite{Banfi:2025mra,Dasgupta:2025cgl}, along with efforts towards their resummation~\cite{Becher:2023mtx,Boer:2023jsy,Boer:2023ljq,Boer:2024hzh,Boer:2024xzy,Becher:2026kbr}.
At the same time, these studies have clarified the role of Glauber gluons in the low-energy logarithmic evolution of hadronic cross sections, showing that they ultimately restore the correct (DGLAP-like) behavior, albeit in a non-trivial and somewhat unexpected way~\cite{Becher:2024kmk,Becher:2025igg}.

Given the intimate connection between collinear-factorization-violating (CFV) terms in scattering amplitudes and their potential impact at the observable level, several studies have been carried out,
both based on amplitude considerations~\cite{Catani:2011st,Cieri:2024ytf,Gardi:2022khw,Gardi:2024axt,Ma:2025emu,Duhr:2025lyg}, as well as on effective field theory frameworks~\cite{Rothstein:2016bsq,Schwartz:2017nmr}.
It was first argued in~\cite{Catani:2011st}, that CFV effects, originating from two-loop amplitudes with one unresolved collinear emission, can arise in dijet cross sections starting at third order ($\mathrm{N^3LO}$) in perturbative QCD.
CFV contributions were found to cancel in the infrared (IR) poles of the cross 
section~\cite{Catani:2011st}, with stronger results established for single-scale 
observables~\cite{Schwartz:2018obd}. More general insights into these cancellation patterns were 
obtained from the study of the soft-collinear behavior of two-loop QCD spacelike splitting 
amplitudes~\cite{Dixon:2019lnw}, and more recently in hard-collinear kinematics in 
$\mathcal{N}=4$ super Yang-Mills (sYM)~\cite{Henn:2024qjq}. Taken together, these results provide strong evidence that CFV terms cancel in two-loop color-summed squared amplitudes, suggesting their absence at the cross-section level at $\mathrm{N^3LO}$ in pure QCD. 

Exploiting recent advances in two-loop five-point QCD amplitudes~\cite{Agarwal:2023suw,DeLaurentis:2023nss,DeLaurentis:2023izi}, in this Letter we  make this picture quantitative. We present two main results: (i) we derive, for the first time, the complete two-loop QCD spacelike splitting amplitudes in all partonic channels, inclusive of all CFV contributions; and (ii) we show that such CFV effects cancel in jet cross sections at $\mathrm{N^3LO}$ QCD. 
These findings provide a decisive step towards ensuring PDF factorization in physical cross sections at this perturbative order.

The letter is organized as follows. We begin by discussing the spacelike splitting kinematics and the origin of factorization-violating phases. 
We then describe two complementary approaches to analyzing the full-color two-loop five-point amplitudes in the spacelike collinear limit, including analytic continuation of the relevant functions, as well as their direct expansion via differential equations methods. From these results, we extract the generalized two-loop splitting amplitudes in QCD, and we show that CFV terms cancel at cross-section level. Furthermore, we
compare our results to the Multi-Regge kinematics (MRK) limit~\cite{Caron-Huot:2020vlo,Buccioni:2024gzo,Abreu:2024xoh} finding interesting connections.
We conclude summarizing our main findings and suggesting future avenues.
%
%
%
\section{Discontinuities and factorization violation }
\label{se:crossing}
Consider an $(n+1)$-point scattering amplitude in QCD, $\mathcal{A}_{n+1}$, where all particles with
momenta $p_i$ are taken to be \emph{outgoing}.
Particles with $p_i^0>0$ (energy) are in the final state, while those with $p_i^0<0$ belong to the initial state.
When a pair of particles $(a,b)$ with helicities $(\lambda_a,\lambda_b$)
become collinear and recombine to form particle $A$ with helicity $\lambda_A$, the amplitude factorizes
as~\cite{Kosower:1999xi}
\begin{equation}
\label{eq:factorizationLP}
\mathcal{A}_{n+1}(\cca{a}^{\lambda_\cca{a}},\ccb{b}^{\lambda_\ccb{b}},\ldots)\simeq
\sum_{\lambda_\cca{A}=\pm} \mathbf{Sp}_{\lambda_{\cca{A}}}(\cca{a}^{\lambda_\cca{a}},\ccb{b}^{\lambda_\ccb{b}})\mathcal{A}_n  (\cca{A}^{\lambda_\cca{A}},\ldots),
\end{equation}
where the relation is valid up to power corrections in the collinear expansion parameter.
Following~\cite{Catani:2011st}, we have introduced a generalized splitting amplitude $\mathbf{Sp}$, which is an operator acting on the color space of the $n$-point sub-amplitude $\mathcal{A}_n$ and it is expanded as
\begin{equation}
\label{eq:perturbativexpansion}
\mathbf{Sp}_{\lambda_\cca{A}}(\cca{a}^{\lambda_\cca{a}},\ccb{b}^{\lambda_\ccb{b}}) = \sum_{\ell=0} \left(\frac{\alpha_s}{4\pi}\right)^\ell
\mathbf{Sp}^{(\ell)}_{\lambda_\cca{A}}(\cca{a}^{\lambda_\cca{a}},\ccb{b}^{\lambda_\ccb{b}}),
\end{equation}
where $\alpha_s$ is the $\overline{\mathrm{MS}}$-renormalized strong coupling constant. 
The scattering amplitudes $\mathcal{A}$ admit the same perturbative expansion.
In Eq.~\eqref{eq:factorizationLP}, the flavor of particle $A$ is fully determined by the pair $(a,b)$. 
In the timelike splitting $(p_a^0 p_b^0>0,\,P^0>0)$
\begin{equation}
\label{eq:timelikesplitt}
    A(P) \to a\left((1-\xi)P\right) + b\left(\xi P\right) \;\;\text{with}\;\;0<\xi<1,
\end{equation}
$\mathbf{Sp}^{(\ell)}$ depends only on the quantum numbers of the collinear pair. In this case, results were presented at one loop in~\cite{Bern:1994zx,Bern:1999ry,Kosower:1999rx}, at two loops in~\cite{Bern:2004cz,Badger:2004uk} with higher-order terms in dimensional regularization given in~\cite{Duhr:2014nda}, and more recently at three-loops in~\cite{Guan:2024hlf}.

In the spacelike case $(p_a^0 p_b^0<0, \,P^0<0)$,
\begin{equation}
\label{eq:spacelikesplitting}
a\left((1-\xi)P\right) \to A(P) + b\left(\xi P\right) \quad\text{with}\quad \xi<0,
\end{equation}
loop corrections induce a non-trivial dependence on the quantum numbers of non-collinear (spectator) partons~\cite{Catani:2011st}. 
This factorization breaking originates from the ambiguity in the sign of the infinitesimal imaginary part of the energy fraction $\xi$, which renders $2 \to n-1$ scattering amplitudes non-analytic in the strict collinear limit and prevents a direct analytic continuation in $\xi$ from the timelike case, Eq.~\eqref{eq:timelikesplitt}.
\begin{figure}[t!]
\includegraphics[scale=1.]{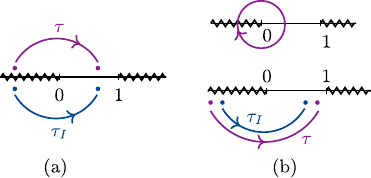}
    \caption{Analytic continuation of $\tau$ $(\equiv \tau_{\cin{\rm in}})$ and $\tau_I$ from the spacelike to the timelike collinear regime; $\cin{\rm in}$ denotes an incoming particle and $I$ an outgoing spectator.
    }
\label{fi:discontinuity}
\end{figure}
To clarify this point, we introduce 
a set of variables $\tau_I$ characterizing the distribution of light-cone energy between parton $a$ and $b$, which all reduce to $\xi$ in collinear limit.  Each $\tau_I$ is formulated in terms of a light-cone direction aligned with a given spectator parton $I$, 
\begin{equation}
\label{eq:ambiguity}
\tau_I = \frac{s_{I\ccb{b}}}{s_{I\ccb{b}}+ s_{I\cca{a}}} \xrightarrow{\cca{a} \parallel \ccb{b}}  |\xi|\,{\rm exp}[- i \pi   (\eta_{I\ccb{b}}- \eta_{I\cca{A}})],
\end{equation}
where $s_{ij} = (p_i+p_j)^2$ and $\eta_{ij} = 1$ if $i,j$ are both outgoing and $\eta_{ij} = 0$ otherwise. For timelike splitting, $\tau_I$ is unambiguously real and positive for all $I$. For spacelike splitting instead, the collinear limit of $\tau_I$ is sensitive to the signature of the spectator momentum, namely $\tau_I \rightarrow -|\xi| \pm i 0^+ $ for incoming/outgoing $I$, 
as illustrated in Fig.~\ref{fi:discontinuity}(a).
 When both incoming and outgoing spectator partons coexist, a direct analytic continuation of the splitting amplitude from the timelike to the spacelike regime is obstructed by an ambiguity originating from the competing $i\pi$ prescriptions of different discontinuity channels. 

In the next section we show how we can overcome this 
hurdle in two alternative ways.
\vspace{-0.3cm}
\section{Spacelike splitting from full-color QCD amplitudes}
\label{se:fivepoint}
In this section, we investigate spacelike factorization at two-loop in full color in QCD, based on the five-point helicity amplitudes of~\cite{Agarwal:2023suw}, where all partonic configurations were presented. These enable a systematic study of all possible quark $(q)$ and gluon $(g)$ configurations of the collinear pair $(\cca{a},\ccb{b})$
in all relevant helicity states.
Starting from the scattering 
kinematics $p_\cin{1} + p_\cca{2} \to p_\ccb{3} + p_4 + p_5$, 
we consider the limit $p_\cca{2} \parallel p_\ccb{3}$,  which amounts to setting  
$\cca{a}=\cca{2}$, $\ccb{b}=\ccb{3}$ in Eq.~\eqref{eq:spacelikesplitting}.

Here, we work with ultraviolet (UV) renormalized, IR divergent five-point scattering amplitudes. The latter are expressed as a linear combination of rational functions $R$ of the Mandelstam invariants $s_{ij}$ and transcendental functions $f$, referred to as massless pentagon functions~\cite{Gehrmann:2018yef,Abreu:2018aqd,Chicherin:2018old,Chicherin:2020oor}, as
\begin{equation}
    \mathcal{A}^{(L)}_5 = \sum_{k=-2L}\sum_m \epsilon^k R^{(k)}_m(s_{ij}) f^{(w_k)}_m(\left\lbrace W \right\rbrace),
\end{equation}
where $\epsilon=(4-D)/2$ is the dimensional regulator, and $w_k\le2L+k$ is the transcendental weight of $f$.
In this work, one-loop amplitudes are expanded up to $k=2$ and two-loop ones up to $k=0$, thus we encounter at most weight-four pentagon functions.
The functions $f$ depend on 31 letters $\left\lbrace W \right\rbrace$ of the pentagon alphabet~\cite{Gehrmann:2018yef,Chicherin:2018old,Chicherin:2020oor}.

To capture the $p_\cca{2}\parallel p_\ccb{3}$ collinear limit, we assign to the invariant $s_{23}$ the scaling $s_{23}\sim \delta^2$, where $\delta$ is a small parameter. A precise parametrization of the kinematics will be given later.
The leading singular behavior of the scattering amplitudes, $\delta^{-1}$, is fully captured by the tree level. Since we work at leading power in $\delta$,
see Eq.~\eqref{eq:factorizationLP}, loop corrections shall only induce logarithmically enhanced terms of the form $\ln^\ell(\delta)$, with $\ell$ being the loop order.

We expand the rational functions $R$ around $\delta=0$ making use of the computer algebra programs \texttt{Form}~\cite{Vermaseren:2000nd,Ruijl:2017dtg}, \texttt{Singular}~\cite{DGPS}, and \texttt{MultivariateApart}~\cite{Heller:2021qkz}. In this procedure, we encounter $\delta^{-3}$ spurious poles
, which force us to Taylor expand the pentagon functions to $\mathcal{O}(\delta^3)$.
In the following, we describe two complementary methods to systematically deal with the transcendental part of the amplitudes.
\vspace{\baselineskip}
\paragraph{\bf{Spacelike to timelike crossing.}}
Based on the discussion of the previous section,
the key observation is that the ambiguity in $\xi$, see Eq.~\eqref{eq:ambiguity}, can be resolved by effectively crossing the spectator $p_{\cin{1}}$ from incoming to outgoing,  bringing $\tau \equiv \tau_{\cin{1}}$ from the upper to lower half complex plane.
For the purpose of analyzing the collinear behavior in scattering amplitudes, this crossing can be efficiently implemented by evaluating the residue around an $S^1$ contour encircling $\tau=0$ in twistor parameters space. Specifically, we employ a modified momentum twistor parametrization \cite{Badger:2013gxa,Abreu:2019rpt} suited for five-point on-shell kinematics in the orientation $p_\cin{1}, p_\ccb{3}, p_\cca{2}, p_4, p_5$, setting
\begin{gather}
x_1 = s\, \tau, \quad x_2  =  s\, c\,  \delta,\quad  x_3 = s\, c\, r \, \delta ,  
 \\
x_4 = \delta \,(1 - \tau), \quad\quad  x_5 = - 1/(c\, \delta) \,.  \nn 
\end{gather}
This framework allows us to take the discontinuity and approach the $p_\cca{2} \parallel p_\ccb{3}$ collinear limit by taking $\delta \to 0$. 
Starting from a generic kinematic point in the spacelike regime,  we access the timelike collinear limit by: 
(1) encircling the branch point at $\tau=0$, (2) taking $\delta \rightarrow 0$, and (3) transitioning from negative to positive real $\tau-$axis. 
The discontinuity evaluated along this full path, depicted in Fig.~\ref{fi:discontinuity}(b), precisely captures the difference between the five-point amplitudes in the spacelike and timelike collinear regimes, thereby determining the CFV terms.

In other words, once we compensate for the discontinuity around $\tau=0$, the five-point amplitude is strictly factorized and governed by the timelike splitting amplitudes which is \emph{analytic} in $\xi$,
\begin{align}\label{eq:discA5Coll}
\mathcal{A}_5   -    [2 \pi i] \,  \text{disc}_{\tau}[\mathcal{A}_5]   \xrightarrow{\cca{2} \parallel \ccb{3}}  {\bf Sp}_{\rm T.L.} (\xi =\tau- i 0^+)  \,  \mathcal{A}_4 
\end{align}
where $ \text{disc}_\tau := \text{disc}^{\circlearrowleft}_{\tau=0}$  is an operator that computes the monodromy around $\tau =0$  following the standard convention $  \text{disc}_{\tau}[\ln \tau]  =  1$.  Acting on the pentagon alphabet letters~\cite{Chicherin:2020oor}, two of them pick up a non-trivial residue
\begin{align}
\text{disc}_\tau[ \ln W_{16}] = 1  , \quad  
 \text{disc}_\tau[ \ln W_{30}] =-1\,.  \nn 
\end{align}
Therefore, to compute $\rm{disc}_{\tau}[\mathcal{A}_5]$, we systematically analyze the discontinuities of the pentagon functions and expand them in the collinear limit, starting at the symbol level.
The result is then Taylor-expanded to the desirable order in $\delta$ by symbol-level manipulations,
\begin{align} 
\mathcal{S}[{\rm disc}_{\tau} F ] = \sum_{n} \delta^n \frac{1}{n!}  \frac{\partial^n}{\partial \delta^n} \mathcal{S}[{\rm disc}_{\tau} F ] \Big|_{\delta=0} \,.
\end{align}
The final step is to promote the symbol-level result to the function-level one, exploiting the 
regularity of the splitting amplitudes in the high-energy regime ($\tau\to\infty$), and the
knowledge of their behavior in Drell-Yan like process~\cite{Badger:2004uk} and in the soft limit~\cite{Dixon:2019lnw}.
In particular, the limit where $\tau \rightarrow 0_{-}$ or  $\tau \rightarrow 1_{+}$ can be fixed by comparing Eq.~\eqref{eq:discA5Coll} against the soft-collinear limit of the five-point amplitudes with $p_\ccb{3} \rightarrow 0$ or $p_\cca{2} \rightarrow 0$ respectively.

Imposing such constraints on $\tau=0_{-},1_{+}$ and $\infty$, we are able to fully determine the five-point amplitudes in the collinear limit at leading power in $\delta$, including the constant $\zeta-$values. 

\vspace{\baselineskip}
\paragraph{\bf{Differential equations method.}}
An alternative approach is to solve the differential equations satisfied by the pentagon functions $f$ with respect to the invariants $s_{ij}$ as a power series in $\delta$. As argued above, we need to obtain up to $\delta^3$ terms in the series. 

We do this following a strategy similar to the one presented in Ref.~\cite{Henn:2024qjq} as well as in Ref.~\cite{Buccioni:2024gzo} for the expansion of five-point amplitudes in the high-energy limit.
We start by parameterizing the kinematics as follows.
\begin{gather}
    s_{12} = s(1+z),\quad s_{23} = -s z \delta^2,\quad s_{34}=s(1-x)z \\
    s_{45} = s, \quad s_{51} = - s \left(1-x - \frac{4y\sqrt{x(1-x)(1+z)}}{1+y^2}\delta\right)\,,\nonumber
\end{gather}
where again the $p_\cca{2}\parallel p_\ccb{3}$ collinear regime is approached by taking $\delta\to0$.
At leading power in $\delta$ 
the 31 letters of the pentagon alphabet~\cite{Chicherin:2018old,Chicherin:2020oor}
reduce to 16 letters organized as~\cite{Henn:2024qjq}
\begin{align}
\label{eq:collinearalphabet}
\nonumber
    &\lbrace \delta\rbrace,\;\, \lbrace s\rbrace, \;\, \lbrace 1\pm y,i\pm y\rbrace,\;\,
    \lbrace x,1-x,z,1+z,1+z-x,\\
    &x+z,1-x-xz,1+z-xz,1+xz,x-z+xz\rbrace\,.
\end{align}

To proceed, we derive the differential equations with respect to the $s_{ij}$ for the complete set of pentagon functions which have been presented as iterated integrals in~\cite{Chicherin:2020oor}.
Then, solving a one-dimensional differential equation in $\delta$, we transport the boundary point of the pentagon functions~\cite{Chicherin:2020oor,Henn:2024qjq}
to $\delta=0$. This enables us to extract the complete tower in $\ln^\ell(\delta)$ 
at leading power in $\delta$. 
Owing to the simplicity of the alphabet Eq.~\eqref{eq:collinearalphabet}, we are able to write a compact ansatz for the functional space in $s$, $y$, and $\lbrace x,z\rbrace$ in terms of generalized polylogarithms (GPLs) up to transcendental weight four. Thanks to the iterated structure of the pentagon functions, we fix the coefficients of the ansatz weight-by-weight by solving the differential equations in these four variables.
The yet undetermined numerical constants at the boundary point can be fixed by performing a high-precision numerical evaluation of the GPLs using \texttt{GiNaC}~\cite{Bauer:2000cp,Vollinga:2004sn} and then analytically reconstruct them in terms of transcendental constants~\cite{Chicherin:2020oor} using the \texttt{PSLQ} algorithm~\cite{Ferguson:PSLQ}.
We use the the \texttt{PolyLogTools}~\cite{Duhr:2019tlz} package for various GPLs manipulations.

Finally, in order to extract the higher terms in $\delta$, we write a generalized power series for $f^{(w)}_i$ as
\begin{equation}
\label{eq:genpowseries}
    f^{(w)}_i = \sum_{n=0}\sum_{\ell=0}^w \delta^n \ln^\ell(\delta) g^{(w-\ell)}_{i,n\ell}(\lbrace s\rbrace,\lbrace y\rbrace,\lbrace x,z\rbrace)\,,
\end{equation}
and solve again the one-dimensional differential equation in $\delta$.
Given the knowledge of the leading power solution, we are able to completely determine the higher-power terms $g_{i,n\ell}$ to arbitrarily high orders.
We have checked Eq.~\eqref{eq:genpowseries} at small values of $\delta$ against numerical evaluations of the pentagon functions in octuple precision using \texttt{PentagonFunctions++}~\cite{Chicherin:2020oor}, finding excellent agreement.

Once rational and transcendental functions in the limit are combined together, remarkably, the scattering amplitude can be expressed in terms of simple single-valued polylogarithms.
Having established agreement between the two methods, we are now in a position to study the full-colour two-loop splitting amplitudes in QCD.

\section{Two-loop generalized splitting amplitudes }
\label{se:generalizedtlspa}
\begin{figure}[t!]
\includegraphics[scale=1.2]{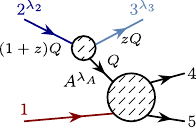}
\caption{Kinematics of the $p_2\parallel p_3$ spacelike collinear splitting}
\label{fi:spacelikesplitting}
\end{figure}

Starting from Eq.~\eqref{eq:factorizationLP}, and redefining $z=-\xi$, $Q=-P$ and $p_\cca{2}=-p_\cca{a}$, 
we consider the spacelike collinear limit $p_\cca{2}\parallel p_\ccb{3}$ depicted in Fig.~\ref{fi:spacelikesplitting}. 
The helicities of particles $a(2)$ and $A$ are in the incoming notation, whereas
$b(3)$ is in the outgoing notation.
Our strategy is to extract the complete expression of $\mathbf{Sp}$ up to two loops, using the results for $\mathcal{A}_5$ in the limit discussed above together with the knowledge of $\mathcal{A}_4$. We take the latter from~\cite{Ahmed:2019qtg,Caola:2021izf,Caola:2022dfa}.
Here we present UV-renormalized amplitudes in ’t~Hooft--Veltman scheme, retaining the full IR structure.

Inspired by Eq.~\eqref{eq:factorizationLP}, we present the main formula of this paper, namely the exponential form of the generalized splitting amplitude
\begin{align}
\label{eq:exponentialform}
    \mathbf{Sp}_{\lambda_\cca{A}}(\cca{2}^{\lambda_\cca{2}},\ccb{3}^{\lambda_\ccb{3}}) &= i\sqrt{4\pi\alpha_s}\,\mathrm{Split}_{\lambda_A}(\cca{2}^{\lambda_\cca{2}},\ccb{3}^{\lambda_\ccb{3}}) \times \\
    & \left\lbrace \mathrm{exp}\left[\mathbf{\boldsymbol{\mathcal{G}}}_{\vec{\lambda}}(z;\epsilon) + \boldsymbol{\Delta}_{\vec{\lambda}}(z_I,\bar{z}_I;\epsilon)\right]\,\mathbf{R}_{\vec{\lambda}}(z)\notag
    \right\rbrace,
\end{align}
where the subscript $\vec{\lambda}=\lbrace\lambda_\cca{A};\lambda_\cca{2},\lambda_\ccb{3}\rbrace$ indicates a dependence on the helicity configuration.
Here, to avoid cluttering the notation, we have suppressed any flavor dependence which is implicitly understood.
In Eq.~\eqref{eq:exponentialform} we introduced the variables~\cite{Henn:2024qjq}
\begin{equation}
    z_I = \frac{\langle \ccb{b}\cca{a}\rangle\langle \cin{\rm in}\,I\rangle}{\langle \cin{\rm in} \,\ccb{b}\rangle\langle \cca{a}I\rangle},\quad \bar{z}_I = z_I^*, \;\;\mathrm{with}\;\; \cca{a}\parallel \ccb{b},
\end{equation}
where $\cin{\mathrm{in}}\neq \cca{a}$ refers to the incoming spectator parton $\cin{1}$ and $I\neq \ccb{b}$ to any outgoing one. 
The color-space operators $\boldsymbol{\mathcal{G}}$, $\boldsymbol{\Delta}$ and $\mathbf{R}$
admit an $\alpha_s$ perturbative expansion as in Eq.~\eqref{eq:perturbativexpansion},
with $\boldsymbol{\mathcal{G}}^{(0)}=\boldsymbol{\Delta}^{(0)}=\boldsymbol{\Delta}^{(1)}=0$\,.
The function $\mathrm{Split}_{\lambda_A}$ captures the leading singular behavior and can be written in terms of spinor products~\cite{Bern:1994zx,Dixon:1996wi}. We collect its explicit expression for each partonic configuration and helicity in the ancillary files as well as in the supplemental material accompanying this publication.
Finally, the operator $\mathbf{R}$, acting on the color space of the hard sub-amplitude, depends on the flavor of the collinear pair $(\cca{a},\ccb{b})$, with $\mathbf{R}^{(0)} \equiv \mathbf{T}^{\cca{c}}_{\cca{a}\ccb{b}}$ and $\mathbf{R}^{(1)}=0$.\footnote{$\mathrm{\mathbf{Sp}}_{\pm}(g^{\mp},g^{\pm})$ presents an exception since it is zero at tree-level, so $\mathbf{R}^{(0)}=0$ and the first non-vanishing operator is $\mathbf{R}^{(1)}\equiv \mathbf{T}^{\cca{c}}_{\cca{a}\ccb{b}}$.}
For $(\cca{a},\ccb{b})=(\cca{2},\ccb{3})$ one has
\begin{equation}
    \mathbf{T}^{\cca{c}}_{\cca{q}\ccb{q}} = T^{\cca{c}}_{\cca{i_2} \ccb{i_3}},\;\;
    \mathbf{T}^c_{\cca{q}\ccb{g}} = T^{\ccb{a_3}}_{\cca{c} \cca{i_2}},\;\;
    \mathbf{T}^c_{\cca{g}\ccb{q}} = T^{\cca{a_2}}_{\ccb{i_3} \cca{c}},\;\;
    \mathbf{T}^c_{\cca{g}\ccb{g}} = i f^{\cca{a_2} \cca{c} \ccb{a_3}},
\end{equation}
where $\cca{c}$ is the color index of particle $\cca{A}$ entering the hard scattering, $\cca{i_2}, \ccb{i_3}$ are indices in the fundamental representation of $SU(N_c)$, and $\cca{a_{2}}, \ccb{a_3}$ are in the adjoint one.

We now describe the various perturbative quantities entering~\eqref{eq:exponentialform}. To improve readability, we drop the explicit dependence on flavor and helicity and reinstate it only when necessary.
$\boldsymbol{\mathcal{G}}^{(n)}$ contains color singlet and dipole contributions, which are \emph{strictly factorized} and \emph{anti-hermitian} respectively. Explicitly,
\begin{equation}
\label{eq:singletdipole}
    \boldsymbol{\mathcal{G}}^{(n)}(z;\epsilon) =  r^{(n)}(z,\epsilon) + 
    \mathcal{C}_{ab}\,\mathbf{T}_\ccb{3}\cdot\mathbf{T}_\cin{1} \bar{\mu}^{n \epsilon} \mathrm{disc}[\bar{r}^{(n)}(z,\epsilon)]\,
\end{equation}
with $\bar{\mu} = \frac{\mu^2}{-s_{ab}}\frac{1+z}{z}$, $\mathcal{C}_{gq}=-2N_c$ and $\mathcal{C}_{ab}=2/N_c$ otherwise.
We obtain the functions $r^{(n)}(z,\epsilon)$ and $\bar{r}^{(n)} \equiv \bar{\mu}^{-n \epsilon} r^{(n)}(z,\epsilon)$ from timelike splitting amplitudes up to two loops in all partonic channels from~\cite{Badger:2004uk}, by taking $w=-z-i0^+$ in their results.
We define $\mathrm{disc}[f(z)] = 2i\mathrm{Im}\left[f(z)\right]$\,.
Explicit results in the spacelike kinematics 
are provided in the supplemental material.

\begin{figure}[t!]
\includegraphics[scale=0.9]{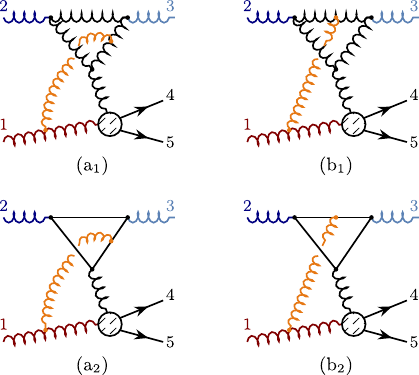}
\caption{Representative planar $(\mathrm{a}_{1,2})$ and non-planar $(\mathrm{b}_{1,2})$ diagrams contributing to the tripole $(\mathrm{a}_{1},\mathrm{b}_{1})$ and $\mathbf{C}_{gg}$ $(\mathrm{a}_{2},\mathrm{b}_{2})$ operators present in Eq.~\eqref{eq:roperator}
.}
\label{fi:r2diagrams}
\end{figure}

The dipole Glauber phase of Eq.~\eqref{eq:singletdipole} is the first source of factorization violation at
amplitude level. At two loops, a second source is given by
\begin{align}
\label{eq:tripoleterm}
 \mathbf{\Delta}^{(2)}_{\vec{\lambda}}  &=   2 i\pi \,\bar{\mu}^{2\epsilon} \sum_{I \in \mathrm{out}}  \mathcal{T}_{\ccb{3}, \cin{1}, I}   \left[  \left(\frac{2}{\epsilon^2} - 2 \zeta_2 \right) (\ln |z_I|^2  + 2i\pi ) \right. \nn\\
 & \left. + 8 \zeta_3  +  h_{\vec\lambda}\,\frac23  \Big( \ln^2 \frac{z_I}{\bar  z_I}  + 4 \pi^2  \Big) \ln \frac{z_I}{\bar  z_I}
 \right]\,,
\end{align} 
where $h_{\vec\lambda}=\pm1$ depends on the helicity configuration. Since their explicit values are irrelevant for the discussion, we provide explicit assignments in the supplemental material.
Here $\mathcal{T}_{\ccb{i}, \cin{j}, k}$ defines the color tripole
\begin{equation}
\label{eq:tripoledefinition}
\mathcal{T}_{\ccb{i}, \cin{j}, k} \equiv [\mathbf{T}_\ccb{i} \cdot \mathbf{T}_\cin{j}, \mathbf{T}_\ccb{i} \cdot \mathbf{T}_{k}]\,. 
\end{equation}
The last source of factorization violation that we observe, which is unique in QCD and it is flavor dependent, is encoded in the operator $\mathbf{R}^{(2)}$. For a collinear pair $(a,b)$ we obtain
\begin{align}
\label{eq:roperator}
    \mathbf{R}^{(2)}_{\cca{a}\ccb{b}} &= (2\pi i) \mathcal{R}_{1,\lambda_A}(\cca{a}^{\lambda_\cca{a}},\ccb{b}^{\lambda_\ccb{b}}) \mathcal{T}_{\ccb{b},\cin{1},\cca{a}} \cdot \mathbf{T}^{\cca{c}}_{\cca{a}\ccb{b}} \nonumber\\
    &+(2\pi i) \mathcal{R}_{2,\lambda_A}(\cca{a}^{\lambda_\cca{a}},\ccb{b}^{\lambda_\ccb{b}}) \, \mathbf{C}^{\cca{c},\cce{e}}_{\cca{a}\ccb{b}}\cdot \mathbf{T}^{\cce{e}}_\cin{1}\,.
\end{align}
The operator $ \mathbf{C}^{\cca{c},e}_{\cca{a}\ccb{b}}$ vanishes when the outgoing collinear parton $b$ is a quark, namely,
$\mathbf{C}^{\cca{c},e}_{\cca{q}\ccb{q}} =  \mathbf{C}^{\cca{c},e}_{\cca{g}\ccb{q}} = 0$,
whereas when $b$ is a gluon we find
\begin{align}
\label{eq:R2gg}
     \mathbf{C}^{\cca{c},\cce{e}}_{\cca{g}\ccb{g}} &= N_f \left[ \mathrm{Tr}\left(T^{\cca{a_2}}T^{\cce{e}}T^{\ccb{a_3}}T^{\cca{c}}\right) + 
     \mathrm{Tr}\left(T^{\cca{c}}T^{\ccb{a_3}}T^{\cce{e}}T^{\cca{a_2}}\right)\right],\\
\label{eq:R2qg}
\mathbf{C}^{\cca{c},\cce{e}}_{\cca{q}\ccb{g}} &= \left(T^d T^{\cce{e}}T^{\ccb{a_3}} T^d\right)_{\cca{c} \cca{i_2}} + 
\frac{1}{N^2_c}\left(T^d T^{[\cce{e}}T^{\ccb{a_3}]} T^d\right)_{\cca{c} \cca{i_2}
},
\end{align}
where $N_f$ is the light-quarks multiplicity. Examples of diagrams contributing to the operator
$\mathbf{R}^{(2)}_{gg}$ are illustrated in Fig.~\ref{fi:r2diagrams}.
The functions $\mathcal{R}_{1,2}$ in Eq.~\eqref{eq:roperator} are IR finite, and they are flavor and helicity dependent.
To highlight their simplicity, here we report
as an example the case of the $g^\pm\to g^\pm g^\pm$ splitting.
We find
\begin{align}
& \mathcal{R}_{1,\pm} (g^\pm, g^\pm) 
  = \frac{4 z \left(11 z^2+21 z+12\right) }{3 (1+z)^3} \left(\text{Li}_2(-z)-\zeta_2 \right)
 \nn \\
& + \frac{22}{3} \ln^2(1 + z) - \frac{4 z(3+2z)}{3(1+z)^2} - \frac{8 z\ln(1 + z)}{3 (1+z)^2}   
\\
& \mathcal{R}_{2,\pm} (g^\pm, g^\pm) = \mathcal{R}_{1,\pm} (g^\pm, g^\pm) \nonumber \\
&+\frac{12z}{1+z}\left(
\zeta_2
- \frac{(1+z)}{2z}\ln^2(1+z)
- \mathrm{Li}_2(-z)
\right).
\end{align}
Remarkably, $\mathcal{R}_{1,2}$ are analytic at $z=0$. They vanish in the soft limit, $z \to 0$,
and in the high-energy limit, $z \to \infty$, they both reduce to constants,
\begin{equation}
\mathcal{R}_1 =  -\frac{8}{3} - \frac{44 \pi^2}{9}, \quad\quad 
\mathcal{R}_2 =  -\frac{8}{3} - \frac{8\pi^2}{9}  \,. 
\end{equation}
Results for the remaining partonic channels and helicity configurations are given in the supplemental material and in the ancillary files. 

As a further check of our results, we compare them with the two-loop MRK limit of~\cite{Buccioni:2024gzo}, focusing on non-power-suppressed amplitudes, namely $g^{\pm}\to g^{\pm}g^{\pm}$ and $q^{\pm}\to g^{\mp} q^{\pm}$.
By suitably rescaling the collinear kinematics to the high-energy limit, and vice versa, the two limits can be taken consecutively.
We find full agreement and, moreover, observe that the operators $\mathbf{\Delta}^{(2)}$ and $\mathbf{R}^{(2)}$ contribute only to a very restricted set of signature amplitudes in MRK, namely those involving a one-loop corrected impact factor with two Reggeon emissions studied in~\cite{Fadin:1999de}. In this context, the operator $\mathbf{C}_{ab}$ contributes only when
such impact-factor correction is associated with the $(2,3)$ projectile.
We leave more in-depth investigations to future work.

In summary, we observe that the generalized splitting
amplitudes, Eq.~\eqref{eq:exponentialform}, take a simple form in QCD.
Up to two-loop order, they can be organized in terms of universal building blocks: an exponentiated Glauber phase comprising of dipole and tripole terms, along with a two-loop IR-finite remainder that emerge in QCD while absent in $\mathcal{N}=4$ sYM. 
The dipole phase in $\boldsymbol{\mathcal{G}}$ can be easily extracted from the discontinuity of two-loop timelike splitting amplitudes~\cite{Bern:2004cz,Badger:2004uk}.
The tripole contribution $\boldsymbol{\Delta}^{(2)}$, which is the same as in $\mathcal{N}=4$ sYM~\cite{Dixon:2019lnw,Henn:2024qjq}, encodes a non-trivial dependence on both \emph{color charge} and \emph{kinematics} of spectator partons, and on the \emph{spin} 
of the collinear partons. 
The two-loop remainder $\mathbf{R}^{(2)}$ is an absorptive term 
describing non-dipole-like Glauber phases. Similarly to the dipole phase, it depends on quantum numbers of the collinear partons and the \emph{color charge} of the incoming spectator.   
Inspecting the color structures of $\mathbf{R}^{(2)}$, we identify that they originate from  loop-induced non-planar corrections which vanish in Drell-Yan-like processes, hence cannot be determined from the timelike splitting amplitudes in Ref.~\cite{Badger:2004uk}.  For $g \to gg$ scattering in particular, the color operators $\mathcal{T}_{\ccb{b},\cin{1},\cca{a}}$ and  $\mathbf{C}^{\cca{c},\cce{e}}_{\cca{g}\ccb{g}}$ can be associated with the non-planar diagrams Fig.~\ref{fi:r2diagrams}$(b_{1})$ and \ref{fi:r2diagrams}$(b_{2})$, respectively, 
each containing a single gluonic or fermionic 
loop. Meanwhile, the planar diagrams Fig.~\ref{fi:r2diagrams}$(a_{1})$
and \ref{fi:r2diagrams}$( a_{2})$ 
contain both dipole and non-dipole-like color components. 
In particular, Fig.~\ref{fi:r2diagrams}$(a_{1})$ (or \ref{fi:r2diagrams}$(a_{2})$) 
can be decomposed in terms of  the color operator $\mathbf{T}_\ccb{3}\cdot\mathbf{T}_\cin{1}$
and $\mathcal{T}_{\ccb{b},\cin{1},\cca{a}}$ 
(or $\mathbf{C}^{\cca{c},\cce{e}}_{\cca{g}\ccb{g}}$). 
Hence both planar and non-planar diagrams in Fig.~\ref{fi:r2diagrams} are expected to contribute to the two-loop IR-finite remainder. 
We note that the $N_f$-dependent
diagrams in Fig.~\ref{fi:r2diagrams} 
do not vanish in an abelian theory (QED), hinting at the presence of non-dipole like CFV effects in the $\cca{2} \parallel \ccb{3}$ collinear limit in the Compton scattering amplitude $e (p_1) \gamma (p_2)  \to  \gamma (p_3) \gamma (p_4) e (p_5)$. 
%
%
%
\section{Cancellation of CFV terms at squared amplitude level}
\label{se:discussion}
Here we discuss the fate of the CFV terms at cross-section level in the collinear limit. For the calculation of cross sections in perturbation theory, one is interested in color-summed squared matrix elements.

At $\mathcal{O}(\alpha_s^2)$, tripole operators acts on tree-level amplitudes and can be shown to vanish once inserted between them and summed over color and spin in pure QCD~\cite{Catani:2011st,Forshaw:2012bi}.
\footnote{
The proof of~\cite{Forshaw:2012bi} relies on tree-level QCD amplitudes being real up to an overall phase. However, for $2\to n$ processes with $n>2$, helicity amplitudes develop non-trivial complex phases. In pure QCD, these phases cancel upon summation over spins.}.
Thus, neither Eq.~\eqref{eq:tripoleterm}, nor the tripole contribution to the operator $\mathbf{R}^{(2)}$ in Eq.~\eqref{eq:roperator} generate CFV terms in the color-summed squared amplitude.

We then focus on the operator in the second line of Eq.~\eqref{eq:roperator}. Upon squaring the matrix element, one gets contributions of the form (neglecting the tripole)
\begin{equation}
\label{eq:roperatorcontraction}
\mathbf{R}^{(0),\dagger}\cdot\mathbf{R}^{(2)} \sim 
(\mathbf{T}^{\cca{\bar{c}}}_{\cca{a}\ccb{b}})^\dagger \cdot \mathbf{C}^{\cca{c},\cce{e}}_{\cca{a}\ccb{b}}\cdot \mathbf{T}^{\cce{e}}_{\cin{1}}\,.
\end{equation}
For $(\cca{a},\ccb{b})=(g,g)$, the operator $\mathbf{T}^{\cca{\bar{c}}}_{\cca{g}\ccb{g}}$ is anti-symmetric in the indices $\cca{a_{2}},\ccb{a_3}$, whereas $\mathbf{C}^{\cca{c},\cce{e}}_{\cca{g}\ccb{g}}$ is symmetric,
see Eq.~\eqref{eq:R2gg}, thus  their contraction 
vanishes upon summing over color indices. As for $(\cca{a},\ccb{b})=(q,g)$, we have instead
\begin{align}
&(\mathbf{T}^{\cca{\bar{c}}}_{\cca{a}\ccb{b}})^\dagger \cdot \mathbf{C}^{\cca{c},\cce{e}}_{\cca{a}\ccb{b}} = \\
&\left(T^d T^{\cce{e}}T^{\ccb{a_3}} T^d T^{\ccb{a_3}}\right)_{\cca{c \bar{c}}} + 
\frac{1}{N^2_c}\left(T^d T^{[\cce{e}}T^{\ccb{a_3}]} T^d T^{\ccb{a_3}}\right)_{\cca{c \bar{c}}
}=0\nonumber\,.
\end{align}

Since both the tripole and $\mathbf{R}^{(2)}$ operators vanish at $\mathcal{O}(\alpha_s^2)$, the dipole term can be written, up to this order, in exponential form. Indeed, it trivially commutes with the singlet operator and can be isolated in a pure phase. 
Therefore, owing to its anti-Hermiticity, the dipole contribution associated with $\mathrm{disc}\left[r^{(L)}(z,\epsilon)\right]$ in Eq.~\eqref{eq:singletdipole}
vanishes in the color-summed squared amplitude.

We therefore establish that all CFV terms arising from two-loop amplitudes with one collinear emission in the spacelike splitting regime cancel at the cross-section level at $\mathrm{N^3LO}$.
%
%
%
\section{Summary and Outlook }
\label{se:conclusion}
In this paper, we have presented, for the first time, the complete expression for two-loop spacelike splitting amplitudes in QCD in full color for arbitrary partonic channels and helicity configurations.
In particular, we provide a full characterization of all collinear factorization-violating terms at this order. 
In contrast to the $\mathcal{N}=4$ sYM analysis~\cite{Henn:2024qjq}, 
we identify new sources of CFV terms at amplitude level at two loops,
Eq.~\eqref{eq:roperator}. All relevant results are provided in the ancillary files accompanying this publication.

As already observed in $\mathcal{N}=4$ sYM~\cite{Henn:2024qjq}, we conclude that at cross-section level, only the color-singlet contribution in Eq.~\eqref{eq:singletdipole} survives, which obeys strict collinear factorization. 
This has important implications for jet cross sections at $\mathrm{N^3LO}$ in QCD: single-unresolved emissions of a collinear parton is independent of the underlying scattering process, and therefore is universal.

A second potential source of collinear factorization violation in $\mathrm{N^3LO}$ cross sections arises from one-loop corrections to double-unresolved emissions, for which results are available in the literature~\cite{Zhu:2020ftr,Czakon:2022dwk,Czakon:2022fqi}. The analysis of Ref.~\cite{Cieri:2024ytf} intriguingly points to the presence of CFV contributions at the amplitude-squared level. This opens the doors to further investigations on the presence of CFV terms in jet observables at $\mathrm{N^3LO}$.

Finally, in the course of this work, we have uncovered interesting connections with the multi-Regge limit of five-point QCD amplitudes. The availability of analytic data opens the possibility of revisiting the infrared collinear regime from a high-energy perspective, and vice versa, offering alternative avenues to explore the interplay between Regge and Glauber dynamics~\cite{Rothstein:2016bsq,Moult:2022lfy,Gao:2024qsg,Rothstein:2024fpx,Gardi:2024axt}. We are excited to pursue this direction further.

{\bf Acknowledgments}
We would like to thank Fabrizio Caola, Lance Dixon and Lorenzo Tancredi for useful discussions and comments on the paper.
We also thank the authors of Ref~\cite{becher_spamp}, in particular Patrick Hager and Dominik Schwienbacher, for detailed comparisons with their independent calculation and for pointing out errors in the supplemental material of the first version.
We thank the Kavli Institute for Theoretical Physics for hosting us during the initial stages of this project.
KY acknowledges CERN TH Department for hospitality while this research was being carried out.
The work of HF and KY is supported by National Natural Science Foundation of China under Grant No. 12357077.

\bibliography{slspa}

\newpage

\onecolumngrid

\section*{Supplemental material}

\makeatletter
\renewcommand\@biblabel[1]{[#1S]}
\makeatother

Here we present explicit results for the collinear splitting in all partonic configurations. 
We define the kinematics of the splitting process as in Fig.~\ref{fi:spacelikesplitting}, namely
\begin{equation}
\label{eq:splittingdefinition}
    a^{\lambda_a}((1+z)Q)\to A^{\lambda_A}(Q) + b^{\lambda_b}(zQ), \quad\quad z>0\,.
\end{equation}
We note that, upon performing the simultaneous replacements $Q\to z Q$ and $z\to (1-z)/z$, one is able to remap the
splitting kinematics to $a(Q) \to A(zQ) + b((1-z)Q)$, with $0<z<1$. 
The helicity labeling respects the incoming convention for particles $a$ and $A$ and
the outgoing one for particle $b$.

For later convenience, we also define the perturbative expansions of the QCD $\beta$-function, the cusp anomalous dimension $\gamma_K$ as well as the quark and gluon collinear anomalous dimensions $\gamma_{q,g}$ as
\begin{equation}
\beta(\alpha_s) = -2\alpha_s\sum_{n=0}\beta_n \left(\frac{\alpha_s}{4\pi}\right)^{n+1},\quad\quad
\gamma_K = \sum_{n=1} \gamma^{(n)}_K \left(\frac{\alpha_s}{4\pi}\right)^n,\quad\quad
\gamma_{q,g} = \sum_{n=1} \gamma^{(n)}_{q,g} \left(\frac{\alpha_s}{4\pi}\right)^n,
\end{equation}
\begin{subequations}
where the relevant coefficients up to two loops are
\begin{align}
\beta_0 &= \frac{11}{3}C_A - \frac{2}{3}N_f, \\
\beta_1 &= \frac{34}{3}C_A^2 - \frac{10}{3}C_A N_f - 2 C_F N_f ,\\
\gamma_K^{(1)} &= 4, \\
\gamma_K^{(2)} &= 8\, C_A\left(\frac{67}{18}-\frac{\pi^2}{6}\right)-\frac{40}{9} N_f, \\
\gamma_q^{(1)} &= -3 C_F ,\\
\gamma_q^{(2)} &= C_F N_f\left(\frac{65}{27} + \frac{\pi^2}{3}\right)
+ C_F^2 \left(-\frac{3}{2} + 2\pi^2 - 24 \zeta_3\right)
+ C_A C_F \left(-\frac{961}{54} - \frac{11\pi^2}{6} + 26 \zeta_3\right), \\
\gamma_g^{(1)} &= -\beta_0 ,\\
\gamma_g^{(2)} &= 2 C_F N_f
+ C_A N_f \left(\frac{128}{27} - \frac{\pi^2}{9}\right)
+ C_A^2 \left(-\frac{692}{27} + \frac{11\pi^2}{18} + 2 \zeta_3\right),
\end{align}
\end{subequations}
with the $SU(N_c)$ Casimir constants $C_A = N_c$ and $C_F = (N_c^2-1)/(2N_c)$\,.
%
%
%
\subsection{\boldmath The $g\to gg$ splitting}
The leading singular behavior $\mathrm{Split}_{\lambda_A}(a^{\lambda_a},b^{\lambda_b})$,
see Eq.~\eqref{eq:exponentialform}, in all helicity configurations is given by
\begin{subequations}
\begin{align}
    & \mathrm{Split}_{+}(g^+,g^+) = \frac{i}{\sqrt{2}\langle ab\rangle}\frac{(1+z)^2}{\sqrt{z(1+z)}},\quad
    &&\mathrm{Split}_{-}(g^-,g^-) = \frac{-i}{\sqrt{2}[ab]}\frac{(1+z)^2}{\sqrt{z(1+z)}},\\
    & \mathrm{Split}_{+}(g^-,g^-) = \frac{i}{\sqrt{2}\langle ab\rangle}\frac{z^2}{\sqrt{z(1+z)}},\quad
    && \mathrm{Split}_{-}(g^+,g^+) = \frac{-i}{\sqrt{2}[ab]}\frac{z^2}{\sqrt{z(1+z)}},\\
    & \mathrm{Split}_{+}(g^+,g^-) = \frac{i}{\sqrt{2}[ab]}\frac{1}{\sqrt{z(1+z)}}, \quad
    &&\mathrm{Split}_{-}(g^-,g^+) =\frac{-i}{\sqrt{2}\langle ab\rangle}\frac{1}{\sqrt{z(1+z)}},\\
    & \mathrm{Split}_{+}(g^-,g^+) = \frac{i [ab]}{\sqrt{2}\langle ab\rangle^2}\frac{1}{\sqrt{z(1+z)}},\quad
    && \mathrm{Split}_{-}(g^+,g^-) = \frac{-i \langle ab\rangle}{\sqrt{2}[ab]^2}\frac{1}{\sqrt{z(1+z)}}\,.
\end{align}
\end{subequations}
We stress once again that the splitting $g^\pm\to g^\mp g^\pm$ vanishes at tree level.
The one-loop coefficients in timelike kinematics are known to all orders in $\epsilon$~\cite{Kosower:1999rx,Badger:2004uk}. Here we report the expression for $r^{(1)}_{\lambda_A}(g^{\lambda_a},g^{\lambda_b})
$, see Eq.~\eqref{eq:singletdipole}, corresponding to the UV-renormalized result in spacelike kinematics in the ’t~Hooft--Veltman scheme.
They read:
\begin{align}
r^{(1)}_{\pm}[g^\pm,g^\mp](z,\epsilon)
&=
c(\epsilon)\left(\frac{\mu^2}{-s_{ab}}\right)^\epsilon
\left\lbrace
\frac{N_c}{\epsilon^2}
\left[
-e^{-i\pi\epsilon}z^\epsilon(1+z)^{-\epsilon}\Gamma(1-\epsilon)\Gamma(1+\epsilon)
+ 2\sum_{n=1}^\infty \epsilon^{2n-1}\mathrm{Li}_{2n-1}\left(\frac{1+z}{z}\right)
\right]
\right. \nonumber \\
&\left.\quad\quad\quad\quad\quad\quad\quad\quad
-\frac{2 z(1+z)}{(1-2\epsilon)(2-2\epsilon)(3-2\epsilon)}
\bigl(N_c(1-\epsilon)-N_f\bigr)
\right\rbrace
-\frac{\beta_0}{2\epsilon},
\label{eq:oneloopggsplit1}
\\
r^{(1)}_{\pm}[g^\pm,g^\pm](z,\epsilon) &=
c(\epsilon)\left(\frac{\mu^2}{-s_{ab}}\right)^\epsilon
\frac{N_c}{\epsilon^2}\left[-e^{-i\pi\epsilon}z^\epsilon(1+z)^{-\epsilon}\Gamma(1-\epsilon)\Gamma(1+\epsilon) + 
2\sum_{n=1}^\infty \epsilon^{2n-1}\mathrm{Li}_{2n-1}\left(\frac{1+z}{z}\right)\right]
-\frac{\beta_0}{2\epsilon},
\label{eq:oneloopggsplit2}
\\
r^{(1)}_{\pm}[g^\mp,g^\pm](z,\epsilon) &=
c(\epsilon)\left(\frac{\mu^2}{-s_{ab}}\right)^\epsilon
\frac{-2 z(1+z)}{(1-2\epsilon)(2-2\epsilon)(3-2\epsilon)}
\bigl(N_c(1-\epsilon)-N_f\bigr)\,,
\label{eq:oneloopggsplit3}
\end{align}
with the additional relations
\begin{equation}
  r^{(1)}_{\pm}[g^\mp,g^\mp](z,\epsilon) =r^{(1)}_{\pm}[g^\pm,g^\pm](z,\epsilon)  \,.
\end{equation}
The quantity $c(\epsilon)$ is defined as 
\begin{equation}
    c(\epsilon) = e^{\epsilon \gamma_E}\frac{\Gamma(1+\epsilon)\Gamma(1-\epsilon)^2}{\Gamma(1-2\epsilon)}\,,
\end{equation}
and $\gamma_E$ is the Euler-Mascheroni constant.
The polylogarithms appearing in Eqs.(\ref{eq:oneloopggsplit1},\ref{eq:oneloopggsplit2}) can be
expressed in terms of real-valued functions using the identity
\begin{equation}
\label{eq:inversionrelation}
    \mathrm{Li}_n\left(\frac{1+z}{z}\right) = (-1)^{n+1}\mathrm{Li}_n\left(\frac{z}{1+z}\right)
    -2\pi i \frac{\left(\ln(1+z)-\ln(z)\right)^{n-1}}{(n-1)!}
    - \frac{(2\pi i)^n}{n!}
    B_n\left(\frac{\ln(1+z)-\ln(z)}{2\pi i}\right),
\end{equation}
where $B_n(x)$ are the Bernoulli polynomials defined in terms of the Bernoulli numbers $B_k$ as
\begin{equation}
    B_n(x) = \sum_{k=0}^n \binom{n}{k}B_k\,x^{n-k}\,.
\end{equation}
The one-loop discontinuity is zero for $r^{(1)}_{\pm}[g^\mp,g^\pm](z,\epsilon)$,
whereas for all other cases is given by
\begin{equation}
\mathrm{disc}[\bar{r}^{(1)}_{\lambda_A}\left[g^{\lambda_a},g^{\lambda_b}](z,\epsilon)\right] = 
-2i\pi N_c \,\frac{c(\epsilon)}{\epsilon}.
\end{equation}
The UV-renormalized two-loop terms contributing to the color-singlet component in Eq.~\eqref{eq:singletdipole} are given by
\begin{align}
\label{eq:ggtwoloop_treeallowed}
r^{(2)}_{\lambda_A}[g^{\lambda_a},g^{\lambda_b}](z,\epsilon)
&= \frac{c(\epsilon)}{c(2\epsilon)}\left(\frac{\beta_0}{\epsilon} + \frac{\gamma^{(2)}_K}{4}\right)\left(r^{(1)}_{\lambda_A}[g^{\lambda_a},g^{\lambda_b}](z,2\epsilon) + \frac{\beta_0}{4\epsilon}\right)\nonumber\\
&\quad- \frac{\beta_0}{\epsilon}r^{(1)}_{\lambda_A}[g^{\lambda_a},g^{\lambda_b}](z,\epsilon) - \frac{\beta_0^2}{4\epsilon^2}-\frac{\beta_1}{4\epsilon}\nonumber\\
&\quad+\frac{c(\epsilon)}{4\epsilon}\left(\frac{\mu^2}{-s_{ab}}\right)^{2\epsilon} e^{2 i \pi \epsilon}z^{-2\epsilon}(1+z)^{-2\epsilon}H_{gg} + \Delta r^{(2)}_{\lambda_A}[g^{\lambda_a},g^{\lambda_b}](z,\epsilon).
\end{align}
This expression holds for $(\lambda_A;\lambda_a,\lambda_b)=(\pm;\pm,\mp)$, $(\pm;\pm,\pm)$ and $(\pm;\mp,\mp)$.
For the configurations that vanish at tree level, we instead find
\begin{align}
r^{(2)}_{\pm}[g^\mp,g^\pm](z,\epsilon) &= -
\left[\frac{3\beta_0}{2\epsilon} + c(\epsilon)\left(\frac{\mu^2}{-s_{ab}}\right)^\epsilon\left(\frac{N_c}{\epsilon^2}e^{i\pi\epsilon}z^{-\epsilon}(1+z)^{-\epsilon} -\frac{\beta_0}{\epsilon}\right)\right]r^{(1)}_{\pm}[g^\mp,g^\pm](z,\epsilon)\nonumber \\
&-\frac{1}{2} \left(r^{(1)}_{\pm}[g^\mp,g^\pm](z,\epsilon)\right)^2+ \Delta r^{(2)}_{\pm}[g^\mp,g^\pm](z,\epsilon),
\end{align}
The IR-finite terms $\Delta r^{(2)}$ are
\begin{align}
\Delta r^{(2)}_{\pm}[g^\pm,g^\mp](z,\epsilon)&= N_c^2 \left(
-\frac{523}{108}
+ \frac{5 \pi^2}{36}
- \frac{11 \pi^4}{720}
\right)
- \frac{z^2 (1+z)^2}{18}  (N_c - N_f)^2
\nonumber \\
&
- z(1+z) \Bigg(
\frac{47 N_c^2}{27}
- \frac{N_f}{2 N_c}
- \frac{23 N_c N_f}{54}
- \frac{22 N_f^2}{27}
+ \frac{1}{3} N_c (N_c - N_f)
\left(
\frac{\pi^2}{3}
+ \frac{\ln(z) - i \pi}{1+z}
- \frac{\ln(1+z)}{z}
\right)
\Bigg)
\nonumber \\
&
+ \beta_0 N_c \Bigg(
-\frac{65}{108}
+ \frac{25 \pi^2}{144}
- \frac{\pi^2}{3} \ln(1+z)
+ (\ln(z) - i \pi)\,\ln^2(1+z)
- \frac{1}{3} \ln^3(1+z)
\nonumber \\
&
+ \ln(1+z)\left(
\frac{\pi^2}{6}
- (\ln(z) - i \pi)\ln(1+z)
- \mathrm{Li}_2(-z)
\right) + (\ln(z) - i \pi)\,\mathrm{Li}_2(-z)
\nonumber \\
&
+ 2\,\mathrm{Li}_3\!\left(\frac{z}{1+z}\right)
+ \frac{55}{12}\,\zeta_3
\Bigg) + \mathcal{O}(\epsilon),
\\
\Delta r^{(2)}_{\pm}[g^\pm,g^\pm](z,\epsilon)&= N_c^2 \Bigg(
-\frac{523}{108}
+ \frac{5 \pi^2}{36}
- \frac{11 \pi^4}{720}
- \frac{3}{2} \bigl(\ln z - i \pi\bigr)
- \frac{3}{2} \ln(1+z)
\Bigg)
\nonumber \\
& + \beta_0 N_c \Bigg(
-\frac{65}{108}
+ \frac{25 \pi^2}{144}
+ \frac{\ln(z)-i \pi}{2}
+ \frac{1}{2} \ln(1+z)
+ \frac{\pi^2}{6} \ln(1+z)
\nonumber \\
&
+ (\ln(z) - i \pi)\mathrm{Li}_2(-z)
- 2 \mathrm{Li}_3(-z)
+ \frac{55}{12} \zeta_3
\Bigg) \nonumber \\
&
+ N_c (N_c - N_f) \Bigg[
\frac{2 z}{3 (1+z)^2} (\ln z - i \pi)\ln(1+z)
- \frac{\ln(z) - i \pi}{3 (1+z)}
+ \frac{2 z}{3 (1+z)^2} \left(\frac{\pi^2}{6} + \mathrm{Li}_2(-z)\right)
\nonumber \\
&
+ \frac{2z(1-z)}{3 (1+z)^3}
\Bigg(
\frac{\ln(z) -i \pi}{2} \left(\frac{\pi^2}{6} - \mathrm{Li}_2(-z)\right)
+ \mathrm{Li}_3(-z) - \zeta_3
\Bigg)
\Bigg]
\nonumber \\
&
+ \beta_0 N_c \frac{z}{1+z}
\Bigg[
(\ln(z) - i \pi)\left(\frac{\pi^2}{6} - \mathrm{Li}_2(-z)\right)
+ 2 \mathrm{Li}_3(-z)
- 2 \zeta_3
\Bigg] + \mathcal{O}(\epsilon),
\\
\Delta r^{(2)}_{\pm}[g^\mp,g^\mp](z,\epsilon)
&= N_c^2\Bigg[
-\frac{523}{108}+\frac{5\pi^2}{36}-\frac{11\pi^4}{720}
-\frac{3}{2}\bigl(\ln z+\ln(1+z)-i\pi\bigr)
\Bigg]\nonumber\\
&+\beta_0 N_c\Bigg[
-\frac{65}{108}+\frac{25\pi^2}{144}+\frac{31}{12}\zeta_3
+\left(\frac{1}{2}+\frac{\pi^2}{6}\right)\bigl(\ln z+\ln(1+z)-i\pi\bigr)
\Bigg]\nonumber\\
&+\frac{2\beta_0 N_c}{z}\Bigg[
\frac{\pi^2}{6}\ln(1+z)+\frac{1}{6}\ln^3(1+z)
+\frac{1}{2}\ln(1+z)\,\mathrm{Li}_2(-z)
-\mathrm{Li}_3(-z)-\mathrm{Li}_3\!\left(\frac{z}{1+z}\right)
\Bigg]\nonumber\\
&+N_c(N_c-N_f)\Bigg[
\frac{2(1+z)}{3z^2}\left(\mathrm{Li}_2(-z)-\frac{\pi^2}{3}\right)
+\frac{\ln(1+z)}{3z}
\Bigg]\nonumber\\
&+N_c(N_c-N_f)\frac{2(1+z)(2+z)}{3z^3}\Bigg[
\frac{\pi^2}{6}\ln(1+z)+\frac{1}{6}\ln^3(1+z)
+\frac{1}{2}\ln(1+z)\,\mathrm{Li}_2(-z)\nonumber\\
&
-\mathrm{Li}_3(-z)-\mathrm{Li}_3\!\left(\frac{z}{1+z}\right)
\Bigg]+\mathcal{O}(\epsilon),
\label{eq:ggfinite_exchanged}\\
\Delta r^{(2)}_{\pm}[g^\mp,g^\pm](z,\epsilon)&=
\Bigg(
4 N_c (N_c - N_f)\Big(
\frac{1}{6}\ln(1+z)
- \frac{z}{6}\big(-i\pi + \ln z - \ln(1+z)\big)
\Big)
\nonumber\\
&+ 4 z(1+z)\Big(
\Big(-\frac{101}{54} + \frac{\zeta_2}{12}
- \frac{\big(-i\pi + \ln z\big)\ln(1+z)}{6}\Big) N_c^2
\nonumber\\
&+ \Big(\frac{451}{216} - \frac{\zeta_2}{12}
+ \frac{\big(\ln z-i\pi\big)\ln(1+z)}{6}\Big) N_c N_f
+ \frac{N_f}{8 N_c}
- \frac{5}{54} N_f^2
\Big)
\Bigg)\,.
\end{align}
and the constant $H_{gg}$~\cite{Badger:2004uk} is expressed as 
\begin{equation}
H_{gg} = \beta_1 + \frac{\pi^2}{16}\,\beta_0\, C_A  \gamma_K^{(1)} + \gamma_g^{(2)}\,.
\end{equation}
Since all functions entering the two-loop finite remainders are real-valued for $z>0$, the corresponding two-loop discontinuities can be readily extracted from the expressions above.
We provide them explicitly in the ancillary files.
The remaining $\mathcal{R}_{1,2}$ introduced in Eq.~\eqref{eq:roperator} are
\begin{align}
\mathcal{R}_{1,\pm}(g^{\pm},g^{\mp}) &= -\frac{4}{3}\Big(
z(3+5z)
+ 2z(1+z)\ln(1+z)
- \frac{11}{2}\ln^2(1+z) \nonumber \\
&\quad\quad\quad\;\;- z(12+3z+2z^2)\big(
\zeta_2 + \mathrm{Li}_2(-z) + \tfrac{1}{2}\ln^2(1+z)
\big)
\Big), \\
\mathcal{R}_{2,\pm}(g^{\pm},g^{\mp}) &= \mathcal{R}_{1,\pm}(g^{\pm},g^{\mp})
- 12z\left(\frac{1+z}{2z}\ln^2(1+z)+\zeta_2 + \mathrm{Li}_2(-z)\right), \\
\mathcal{R}_{1,\pm}(g^{\mp},g^{\mp}) &= -\frac{8}{3}(1+11\zeta_2), \qquad
\mathcal{R}_{2,\pm}(g^{\mp},g^{\mp}) = -\frac{8}{3}(1+2\zeta_2),
\end{align}
and for the helicity configuration which vanishes at tree level
\begin{equation}
    \mathcal{R}_{1,\pm}(g^{\mp},g^{\pm})= -4z^2,\quad\quad\quad
    \mathcal{R}_{2,\pm}(g^{\mp},g^{\pm})= -4z^2\,.
\end{equation}
Finally, for the helicity-dependent sign in Eq.~\eqref{eq:tripoleterm} we find $h_{\vec{\lambda}} = \lambda_b$, except for $(\lambda_A;\lambda_a,\lambda_b)=(\pm;\mp,\mp)$, where $h_{\vec{\lambda}} = -\lambda_b$.
%
\subsection{\boldmath The $q\to qg$ splitting}
The leading singular behavior of the non-vanishing splitting processes is given by
\begin{subequations}
\begin{align}
& \mathrm{Split}_{-}(q^-,g^-) = \frac{i\sqrt{2}}{\sqrt{z}[ab]},
&&\mathrm{Split}_{+}(q^+,g^+) = \frac{-i\sqrt{2}}{\sqrt{z}\langle ab\rangle},\\
& \mathrm{Split}_{-}(q^-,g^+) = \frac{i\sqrt{2}(1+z)}{\sqrt{z}\langle ab\rangle},
&&\mathrm{Split}_{+}(q^+,g^-) = \frac{-i\sqrt{2}(1+z)}{\sqrt{z}[ab]}\,.
\end{align}
\end{subequations}
The equilavent factors for $\bar{q}\to\bar{q}g$ can be simply obtained via charge-conjugation of the spinor products, see e.g~\cite{Dixon:1996wi}.
The one-loop UV-renormalized results in spacelike kinematics are
\begin{align}
r^{(1)}_{\pm}[q^\pm,g^\pm](z,\epsilon) &= -c(\epsilon)\left(\frac{\mu^2}{-s_{ab}}\right)^\epsilon\bigg\lbrace\frac{N_c}{\epsilon^2}
\left[
1-\sum_{n=1}^\infty \epsilon^n\left(\mathrm{Li}_{n}\left(\frac{1+z}{z}\right)
-\frac{1}{N^2_c}\mathrm{Li}_{n}\left(\frac{z}{1+z}\right)
\right)
\right]\nonumber \\
&\quad\quad\quad\quad\quad\quad\quad\quad\;+ \left(\frac{N_c^2+1}{N_c}\right)\frac{z}{2(1-2\epsilon)}
\bigg\rbrace -\frac{\beta_0}{2\epsilon}, \\
r^{(1)}_{\pm}[q^\pm,g^\mp](z,\epsilon) &= -c(\epsilon)\left(\frac{\mu^2}{-s_{ab}}\right)^\epsilon\bigg\lbrace\frac{N_c}{\epsilon^2}
\left[
1-\sum_{n=1}^\infty \epsilon^n\left(\mathrm{Li}_{n}\left(\frac{1+z}{z}\right)
-\frac{1}{N^2_c}\mathrm{Li}_{n}\left(\frac{z}{1+z}\right)
\right)
\right]
\bigg\rbrace -\frac{\beta_0}{2\epsilon}\,,
\end{align}
where Eq.~\eqref{eq:inversionrelation} can be used to extract explicitly the imaginary terms in the prescription $z\to z+i0^{+}$\,.
The one-loop discontinuity is the same as in the $g\to gg$ splitting and it reads
\begin{equation}
\label{eq:oneloopdiscres}
\mathrm{disc}[\bar{r}^{(1)}_{\lambda_A}[q^{\lambda_a},g^{\lambda_b}](z,\epsilon)] = 
-2i\pi N_c \,\frac{c(\epsilon)}{\epsilon}.
\end{equation}
The UV-renormalized two-loop terms contributing to the color-singlet component in Eq.~\eqref{eq:singletdipole}
are organized as
\begin{align}
r^{(2)}_{\lambda_A}[q^{\lambda_a},g^{\lambda_b}](z,\epsilon) &= \frac{c(\epsilon)}{c(2\epsilon)}\left(\frac{\beta_0}{\epsilon} + \frac{\gamma^{(2)}_K}{4}\right)\left(r^{(1)}_{\pm}[q^{\lambda_a},g^{\lambda_b}](z,2\epsilon) + \frac{\beta_0}{4\epsilon}\right)
- \frac{\beta_0}{\epsilon}r^{(1)}_{\pm}[q^{\lambda_a},g^{\lambda_b}](z,\epsilon) - \frac{\beta_0^2}{4\epsilon^2}-\frac{\beta_1}{4\epsilon} \nonumber \\
&+\frac{c(\epsilon)}{4\epsilon}\left(\frac{\mu^2}{-s_{ab}}\right)^{2\epsilon} e^{2 i \pi \epsilon}z^{-2\epsilon}(1+z)^{-2\epsilon}H_{qg} + \Delta r^{(2)}_{\pm}[q^{\lambda_a},g^{\lambda_b}](z,\epsilon),
\end{align}
where the constant $H_{qg}$ is expressed as 
\begin{equation}    
H_{qg} = \beta_1 + \frac{\pi^2}{16}\,\beta_0\, C_A  \gamma_K^{(1)} -\frac{1}{4}\beta_0\gamma_K^{(2)} -\frac{1}{4}\gamma^{(1)}_g\gamma_K^{(2)} + \gamma_g^{(2)}\,.
\end{equation}
Since the expressions for the finite remainder are rather lengthy, we refrain from presenting them here. Their complete form, as well as the their discontinuities are provided in the ancillary files.
The functions $\mathcal{R}_{1,2}$ of Eq.~\eqref{eq:roperator} in this case are as follows
\begin{subequations}
\begin{align}
\mathcal{R}_{1,\pm}(q^{\pm},g^{\pm}) &= -6z
+ \frac{\pi^2}{3} z(4+z)
+ 2(1-z)\ln(1+z)
+ (3+4z+z^2)\ln^2(1+z)
- \frac{8}{3}\ln^3(1+z) \nonumber \\
&+ \big(2z(4+z) - 8\ln(1+z)\big)\mathrm{Li}_2(-z)
+ 8\,\mathrm{Li}_3(-z)
+ 8\,\mathrm{Li}_3\!\left(\frac{z}{1+z}\right),
\\
\mathcal{R}_{2,\pm}(q^{\pm},g^{\pm}) &= \mathcal{R}_{1,\pm}(q^{\pm},g^{\pm})
-12z\left(
\zeta_2
+ \frac{1+z}{2z}\ln^2(1+z)
+ \mathrm{Li}_2(-z)
\right),
\\
\mathcal{R}_{1,\pm}(q^{\pm},g^{\mp}) &=
-\frac{6z}{1+z}
+ \frac{2\ln(1+z)}{1+z}
+ 3\ln^2(1+z)
- \frac{2z(4+3z)}{(1+z)^2}\left(\frac{\pi^2}{6}-\mathrm{Li}_2(-z)\right)\nonumber \\
&- 8\left(
\frac{1}{3}\ln^3(1+z)
+ \ln(1+z)\,\mathrm{Li}_2(-z)
- \mathrm{Li}_3(-z)
- \mathrm{Li}_3\!\left(\frac{z}{1+z}\right)
\right),
\\
\mathcal{R}_{2,\pm}(q^{\pm},g^{\mp}) &= \mathcal{R}_{1,\pm}(q^{\pm},g^{\mp}) + \frac{12z}{1+z}\left(
\zeta_2
- \frac{1+z}{2z}\ln^2(1+z)
- \mathrm{Li}_2(-z)
\right)\,.
\end{align}
\end{subequations}
The helicity-dependent sign in Eq.~\eqref{eq:tripoleterm} is given by $h_{\vec{\lambda}} = \lambda_b.$
Equivalent loop-level results for the $\bar{q}\to\bar{q}g$ splitting can be obtained from the above ones,
via e.g.~ $r^{(n)}_{\lambda_A}[\bar{q}^{\lambda_a},g^{\lambda_b}] = r^{(n)}_{\lambda_A}[q^{-\lambda_a},g^{\lambda_b}]$\,.
\subsection{\boldmath The $g\to \bar{q}q$ splitting}
The leading singular behavior of the non-vanishing splitting processes is given by
\begin{subequations}
\begin{align}
& \mathrm{Split}_{+}(g^+,q^-) = \frac{-i\sqrt{2}}{\sqrt{1+z}[ab]},
&&\mathrm{Split}_{-}(g^-,q^+) = \frac{i\sqrt{2}}{\sqrt{1+z}\langle ab\rangle},\\
& \mathrm{Split}_{+}(g^-,q^-) = \frac{-i\sqrt{2}z}{\sqrt{1+z}\langle ab\rangle},
&&\mathrm{Split}_{-}(g^+,q^+) = \frac{ i\sqrt{2}z}{\sqrt{1+z}[ab]}\,.
\end{align}
\end{subequations}
The one-loop UV-renormalized results in spacelike kinematics are given by
\begin{align}
r^{(1)}_{\pm}[g^\pm,q^\mp](z,\epsilon) &= -c(\epsilon)\left(\frac{\mu^2}{-s_{ab}}\right)^\epsilon\bigg\lbrace\frac{N_c}{\epsilon^2}
\left[
1-\sum_{n=1}^\infty \epsilon^n\left(\mathrm{Li}_{n}\left(\frac{z}{1+z}\right)
-\frac{1}{N^2_c}\mathrm{Li}_{n}\left(\frac{1+z}{z}\right)
\right)
\right]\nonumber \\
&\quad\quad\quad\quad\quad\quad\quad\quad\;- \left(\frac{N_c^2+1}{N_c}\right)\frac{1+z}{2(1-2\epsilon)}
\bigg\rbrace -\frac{\beta_0}{2\epsilon}, \\
r^{(1)}_{\pm}[g^\mp,q^\mp](z,\epsilon) &= -c(\epsilon)\left(\frac{\mu^2}{-s_{ab}}\right)^\epsilon\bigg\lbrace\frac{N_c}{\epsilon^2}
\left[
1-\sum_{n=1}^\infty \epsilon^n\left(\mathrm{Li}_{n}\left(\frac{z}{1+z}\right)
-\frac{1}{N^2_c}\mathrm{Li}_{n}\left(\frac{1+z}{z}\right)
\right)
\right]
\bigg\rbrace -\frac{\beta_0}{2\epsilon}\,,
\end{align}
and the one-loop discontinuity entering Eq.~\eqref{eq:singletdipole} is
\begin{equation}
\mathrm{disc}[\bar{r}^{(1)}_{\lambda_A}[g^{\lambda_a},q^{\lambda_b}](z,\epsilon)] = 
\frac{2\pi i}{N_c} \,\frac{c(\epsilon)}{\epsilon}.
\end{equation}
Similarly to the previous sections, the two-loop results are organized as
\begin{align}
r^{(2)}_{\lambda_A}[g^{\lambda_a},q^{\lambda_b}](z,\epsilon) &= \frac{c(\epsilon)}{c(2\epsilon)}\left(\frac{\beta_0}{\epsilon} + \frac{\gamma^{(2)}_K}{4}\right)\left(r^{(1)}_{\pm}[g^{\lambda_a},q^{\lambda_b}](z,2\epsilon) + \frac{\beta_0}{4\epsilon}\right)
- \frac{\beta_0}{\epsilon}r^{(1)}_{\pm}[g^{\lambda_a},q^{\lambda_b}](z,\epsilon) - \frac{\beta_0^2}{4\epsilon^2}-\frac{\beta_1}{4\epsilon} \nonumber \\
&+\frac{c(\epsilon)}{4\epsilon}\left(\frac{\mu^2}{-s_{ab}}\right)^{2\epsilon} e^{2 i \pi \epsilon}z^{-2\epsilon}(1+z)^{-2\epsilon}H_{gq} + \Delta r^{(2)}_{\pm}[g^{\lambda_a},q^{\lambda_b}](z,\epsilon),
\end{align}
with $H_{gq} = H_{qg}$. The complete expressions for the finite remainders, together with the two-loop discontinuities $\mathrm{disc}[\bar{r}^{(2)}(z,\epsilon)]$, are provided in the ancillary files.
Since the operator $\mathbf{C}_{gq}$ vanishes, we only report the function $\mathcal{R}_{1}$, 
\begin{subequations}
\begin{align}
\mathcal{R}_{1,\pm}(g^{\pm},q^{\mp}) &= 12z\left(
\zeta_2
+ \frac{(1+z)\,\ln^2(1+z)}{2z}
+ \mathrm{Li}_2(-z)
\right),\\
\mathcal{R}_{1,\pm}(g^{\mp},q^{\mp}) &= -4\pi^2\,.
\end{align}
\end{subequations}
The helicity-dependent sign in Eq.~\eqref{eq:tripoleterm} is assigned as $h_{\vec{\lambda}} = -\lambda_a.$
%
%
%
\subsection{\boldmath The $q\to gq$ splitting}
%
%
The final splitting process we consider is when an initial-state quark splits 
into an outgoing quark and a gluon which enters the hard process.
The independent leading-order factors are
\begin{subequations}
\begin{align}
&\mathrm{Split}_{+}(q^+,q^+) = \frac{-i\sqrt{2}\,z}{\langle ab\rangle},
&& \mathrm{Split}_{-}(q^-,q^-) = \frac{i\sqrt{2}\,z}{[ab]},\\
&\mathrm{Split}_{-}(q^+,q^+) = \frac{i\sqrt{2}\,(1+z)}{[ab]},
&& \mathrm{Split}_{+}(q^-,q^-) = \frac{-i\sqrt{2}\,(1+z)}{\langle ab\rangle},
\end{align}
\end{subequations}
and the remaining configurations can be obtained by means of charge and parity conjugation operations~\cite{Badger:2004uk}.
The only independent UV-renormalized one-loop singlet contribution reads
\begin{align}
r^{(1)}_{\pm}[q^\pm,q^\pm](z,\epsilon)
&=
c(\epsilon)\left(\frac{\mu^2}{-s_{ab}}\right)^\epsilon
\left\lbrace
\frac{N_c}{\epsilon^2}
\left[
1-e^{-i\pi\epsilon}z^\epsilon(1+z)^{-\epsilon}\Gamma(1-\epsilon)\Gamma(1+\epsilon)
+ 2\sum_{n=1}^\infty \epsilon^{2n-1}\mathrm{Li}_{2n-1}\left(\frac{1+z}{z}\right)
\right]
\right. \nonumber \\
&\left.\quad\quad
+N_c\left(\frac{13 - 8\epsilon}{2\epsilon(3 - 2\epsilon)(1 - 2\epsilon)}\right)
+ \frac{1}{N_c}\left(\frac{2 - \epsilon + 2\epsilon^2}{2\epsilon^2(1 - 2\epsilon)}\right)
- 2N_f\left(\frac{1 - \epsilon}{\epsilon(3 - 2\epsilon)(1 - 2\epsilon)}\right)
\right\rbrace
-\frac{\beta_0}{2\epsilon}\,,
\end{align}
while the discontinuity in the dipole term is given by
\begin{equation}
\mathrm{disc}\left[\bar{r}^{(1)}_{\lambda_A}[q^{\lambda_a},q^{\lambda_b}](z,\epsilon)\right] = 
-2i\pi N_c \,\frac{c(\epsilon)}{\epsilon}.
\end{equation}
The two-loop correction to the singlet component reads
\begin{align}
r^{(2)}_{\lambda_A}[q^{\lambda_a},q^{\lambda_b}](z,\epsilon) &= \frac{c(\epsilon)}{c(2\epsilon)}\left(\frac{\beta_0}{\epsilon} + \frac{\gamma^{(2)}_K}{4}\right)\left(r^{(1)}_{\lambda_A}[q^{\lambda_a},q^{\lambda_b}](z,2\epsilon) + \frac{\beta_0}{4\epsilon}\right)
- \frac{\beta_0}{\epsilon}r^{(1)}_{\lambda_A}[q^{\lambda_a},q^{\lambda_b}](z,\epsilon) - \frac{\beta_0^2}{4\epsilon^2}-\frac{\beta_1}{4\epsilon} \nonumber \\
&+\frac{c(\epsilon)}{4\epsilon}\left(\frac{\mu^2}{-s_{ab}}\right)^{2\epsilon} e^{2 i \pi \epsilon}z^{-2\epsilon}(1+z)^{-2\epsilon}H_{qq} + \Delta r^{(2)}_{\lambda_A}[q^{\lambda_a},q^{\lambda_b}](z,\epsilon),
\end{align}
where we introduced the constant term
\begin{align}
H_{qq} = \beta_1
- \gamma_g^{(2)}
- \frac{1}{16} \beta_0 C_A \pi^2 \gamma_K^{(1)}
+ \frac{1}{8} \beta_0 C_F \pi^2 \gamma_K^{(1)}
- \frac{1}{4} \beta_0 \gamma_K^{(2)}
+ \frac{1}{4} \gamma_g^{(1)} \gamma_K^{(2)}
- \frac{1}{2} \gamma_K^{(2)} \gamma_q^{(1)}
+ 2 \gamma_q^{(2)}\,. 
\end{align}
The two-loop finite remainders and discontinuities for both helicity classes $(\pm;\pm\pm)$ and $(\mp;\pm\pm)$ are given in the ancillary files.
As in the $g\to\bar{q}q$ splitting, the operator $\mathbf{C}_{qq}$ vanishes. For the two helicity classes we find
\begin{align}
    \mathcal{R}_{1,\pm}(q^\pm,q^\pm) &= -4\pi^2\,,\\
    \mathcal{R}_{1,\mp}(q^\pm,q^\pm) &= 6\ln^2(1+z) + \frac{12z}{1+z}\left[\mathrm{Li}_2(-z)-\zeta_2\right].
\end{align}
Finally, for $h_{\vec{\lambda}}$ in Eq.~\eqref{eq:tripoleterm}, $h_{\vec{\lambda}} = \lambda_A.$
\end{document}